\def\apj{{ApJ}}

\def\mnras{{MNRAS}}

\def\bh{black hole~}

\def\ers{{\rm erg/sec}}
\def\ms{M_{\odot}}
\def\et{et al.\ }
\def\rev{reverberation~}
\def\vFWHM{\ifmmode v_{\mbox{\tiny FWHM}} \else
            $v_{\mbox{\tiny FWHM}}$\fi}
\def\kms{\ifmmode {\rm km\ s}^{-1} \else km s$^{-1}$\fi}
\def\ers{\ifmmode {\rm erg\ s}^{-1} \else erg s$^{-1}$\fi}

\documentclass[cup5b]{caps}
\usepackage{graphicx}
\usepackage{amssymb}
\usepackage{ociwsymp1e}
\HeadText{A. Wandel}

\begin{document}

\pagenumbering{arabic}

\author[]{A. WANDEL \\ The Hebrew University of Jerusalem
}

\chapter{Relations between Massive Black \\
         Holes in AGN and their Host \\ Galaxies}

\begin{abstract}

Massive black holes detected in the centers of many nearby galaxies are linearly correlated
 with the luminosity of the host bulge, the
black hole mass being about 0.1\% of the bulge mass. 
An even stronger relation exists between the BH  mass (Mbh) and the stellar velocity 
dispersion in the host bulge.
We show that massive BHs of
AGNs (measured by reverberation mapping) and their bulge luminosity 
( measured by  using a bulge/disk decomposition)
follow the same relations as ordinary (inactive) galaxies, with the exeption of 
narrow line AGN which apparently have significantly lower BH/bulge ratios.
Narrow line AGNs seem to be outstanding also in the Mbh-velocity dispersion relation: 
the small number of
Seyfert galaxies with measured velocity dispersion indicate that narrow line Seyfert 1 galaxies
have a smaller BH mass/velocity dispersion than quiescent galaxies and broad line Syferts. 
Estimating the velocity dispersion for 
from the bulge luminosity with the Faber-Jackson relation
more than doubles the sample and supports these results.

\end{abstract}

\section{Introduction}

Over the past decade there has been a dramatic confirmation of the
basic paradigm in which AGN are energized by the accretion of
matter onto massive black holes (MBHs). 
%
Massive Black Holes (MBHs) detected in the centers of many nearby non-active galaxies 
(Kormendy and Richstone 1995, Kormendy \& Gebhardt (2002))
show an 
approximately linear relation with the luminosity of the host bulge, inferring the black 
hole mass is 0.001-0.002 of the bulge. 
In addition to the MBHs detected by techniques of stellar and gas kinematics,
the masses of about three dozen  MBHs in AGNs have been estimated  by reverberation 
mapping of the broad emission-line region. 
High quality reverberation data and virial BH mass estimates are presently available 
for 20 Seyfert 1 nuclei (Wandel, Peterson and Malkan 1999, hereafter WPM) 
and 17 PG quasars (Kaspi \et 2000).
The virial  estimate has been shown to be consistent with the real BH mass 
(Peterson \& Wandel 1999; 2000).
Previous work suggested that MBHs of active galactic nuclei 
(Seyfert galaxies and quasars) follow a similar relation (Laor 1998; Ho 1999; Wandel 1999). 
New and updated data for AGN confirm that
AGN and  quiescent galaxies have the same BH-bulge mass relation
(Wandel 2002; McLure \& Dunlop 2003). 
A strong and tight  relation was found between the BH mass and the central stellar 
velocity dispersion in the host galaxy:
Ferrarese and Merritt (2000) and Gebhardt \et (2000a) have found that MBHs of inactive
galaxies is better correlated with the stellar velocity dispersion in the bulge 
than with the bulge luminosity. Apparently this relation holds also for AGNs:
the few Seyfert galaxies with stellar velocity data and reverberation BH mass
estimates seem to be consistent with the BH-velocity dispersion relation of inactive galaxies
(Gebhardt \et 2000b), a conclusion strengthened by recent observations of the
velocity dispersion in  \rev mapped Seyfert galaxies (Ferrarese \et 2001). 
\section{Results}
\begin{figure}
    \centering
\includegraphics[height=90mm,width=145mm]{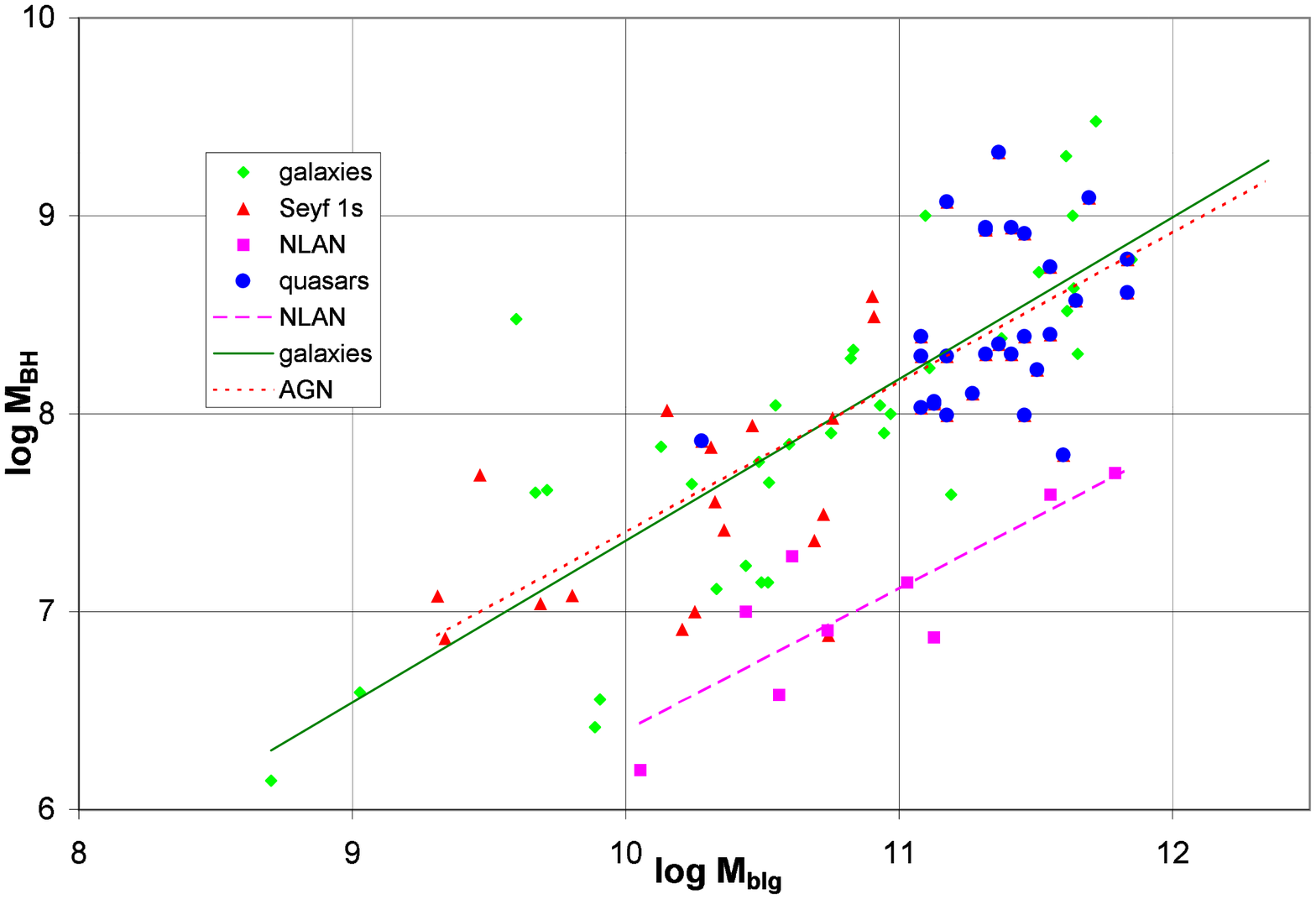}
  \caption {Black hole mass vs. bulge mass. Green diamonds represent quiescent galaxies,
red triangles and blue circles are broad line Seyferts and quasars, respectively, pink 
squares are narrow line AGNs}
   \label{fig1}
  \end{figure}
We combine new and updated data to yield a data base of  55 AGNs (28 quasars,
18 Seyfert 1 galaxies and 9 narrow line AGNs).
Most of these objects have \rev mapping masses and 
almost all of them have
the bulge luminosity measured directly by using  bulge/disk decomposition
(McLure \& Dunlop 2000).
Our results  (figure \ref{fig1}) indicate that broad-line
AGNs (Wandel 2002) follow the same BH-bulge relation as ordinary (inactive) 
galaxies ( Kormendy and  Gebhardt 2002), while narrow line AGNs have significantly lower BH/bulge
mass ratios.
We find that broad line AGNs have an average black hole/bulge mass fraction of 
$\sim 0.0015$ with a strong correlation ($M_{\rm BH}\propto L_{\rm bulge}^{0.95\pm 0.16}$). 
This BH-bulge relation is consistent with the BH-bulge relation of quiescent galaxies and 
 tighter than previous results.
We define narrow line AGNs as having relatively narrow FWHM regardless of their 
other properties (e.g. X-ray spectra). However,  the limit between narrow and broad
line objects is somewhat arbitrary, and as we show below, the results do not really depend on
where one places the border line. Also, some of the narrow
line objects may have a poorly resolved broad component, hence not being intrinsically narrow.
 Narrow line AGNs (Narrow Line Seyfert 1s and quasars, defined here as AGNs having permitted lines 
narrower than 2200{\rm km/s} ) appear to have a lower BH/bulge ratio, 
$M_{\rm BH}/M_{\rm bulge}\sim 10^{-4}-10^{-3}$.
A similar suggestion was made by Mathur et al (2001) on the basis of BH masses estimated mainly from
accretion disk fitting.
It is not obvious that 
the lower BH/bulge ratio of narrow line AGNs reflects also in the $\sigma-M_{BH}$ relation:
 the few narrow line Seyfert 1 galaxies with measured stellar velocity dispersion seem to 
be consistent with $\sigma-M_{BH}$
ratio as broad line AGNs and quiescent galaxies. We use the Faber-Jackson relation to
predict the velocity dispersion of AGNs and combine it with the measured \rev BH masses 
to give the $\sigma-M_{BH}$ relation for over 50 AGNs. 
 
\section{Dependence on the Broad Emission-Line Width} 
The lower BH/bulge ratios of narrow-line AGN  seems to be part of a continous trend:
we find that the $M_{\rm BH}/L_{\rm bulge}$ ratio in AGN is strongly correlated with
the emission-line width, 
$M_{\rm BH}/L_{\rm bulge}\sim FWHM(H\beta)^2$
\begin{figure}
    \centering
\includegraphics[height=80mm,width=125mm]{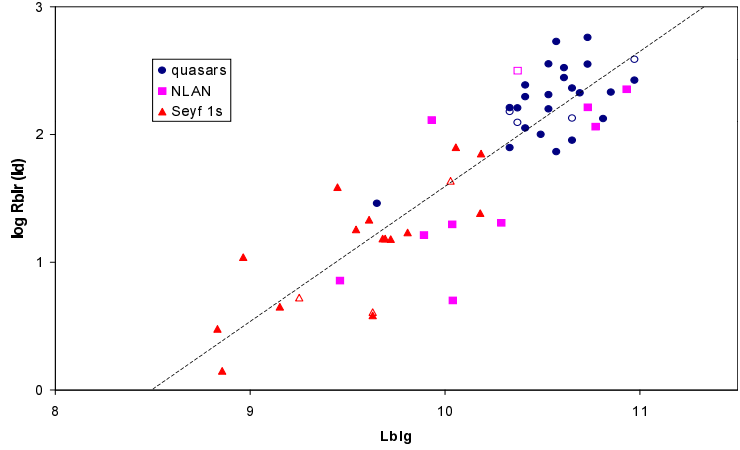}
  \caption {Size of broad emission-line region (BLR) vs. bulge luminosity. 
Triangles and circles are broad line Seyferts and quasars, respectively, squares are narrow line AGNs}
 \label{fig2}
 \end{figure}

This tight correlation is not surprising - it is probably reflecting the virial relation
$M_{\rm BH}=1.46\times 10^5 \ms\left ({R_{blr}\over {\rm lt-days}}\right ) 
\left ({v_{FWHM}\over 10^3\kms }\right )^2$.
Combining the two relations and cancelling out the dependence on the
line width (velocity) on both sides of the equation we find a new, independent relation:
$$R_{BLR}\approx 27 \left ( {L_{\rm bulge}\over 10^{10} 
L_\odot}\right )\quad {\rm lt-days}.$$ 

a very tight correlation between bulge luminosity and BLR size (fig.\ref {fig2})
The best fit to all 55 AGNs in our sample  is
$R= 13.5 L_{10}^{1.05}\quad {\rm lt-days} $
with a correlation coefficient of 0.91.

This is a non trivial
correlation, as it relates two independent observables: on the one hand
the bulge luminosity, a global galactic property on a scale of kpc,
and on the other hand the distance of the line-emitting ionized gas
from the central continuum source - the broad  line region (BLR) size 
measured by reverberation mapping - on
a scale of few light days.
The correlation may be caused by the BLR radius scaling with the AGN luminosity
via the Eddington ratio, but it is
much stronger than the BH-bulge relation which supports the possibility of this
new relation being more fundamental.
Interestingly, Narrow line AGNs are not different from broad line AGNs  in this new relation.

\section{The BH Mass - Velocity Dispersion Relation in AGN}

We find that in AGNs with "narrow" boad lines (e.g. H$\beta$)  the \bh /bulge ratio
 seems to be significantly smaller 
than in broad line AGNs and in normal galaxies.
Yet, in earlier works Seyfert galaxies seem to be consistent with the $\sigma -M_{\rm BH}$ relation of inactive 
galaxies (Gebhardt \et 2000b; Merritt \& Ferrarese 2001b). 
The question whether NLANs have smaller BH masses can therefore be approached by looking 
more carefully at their 
location in the $\sigma -M_{\rm BH}$ plane.
The bulge velocity dispersion is measured for only 11 Seyfert 1 galaxies and one radio galaxy.
Of these 4 can be classified as having narrow lines (Mrk 110, Mrk 590, 3C120 and NGC 4051) and 8
as broad-line Seyferts.

\begin{figure}
    \centering
\includegraphics[height=90mm,width=145mm]{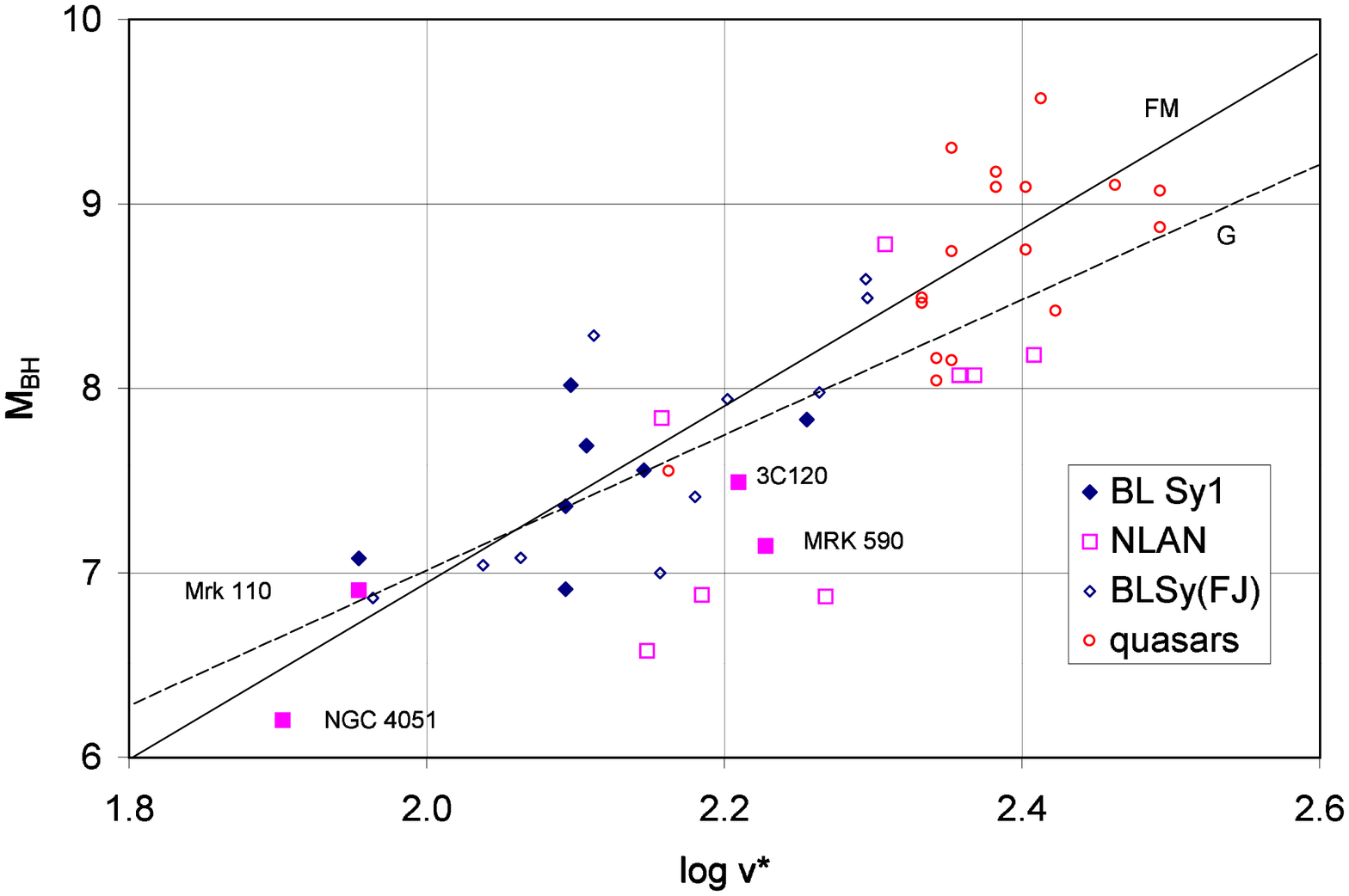}
\caption{Black hole mass of AGNs plotted against the stellar velocity dispersion ($v^*=\sigma$).
Blue triangles are broad line Seyferts,  pink 
squares denote NLS1s and red 
circles denote quasars.
Solid symbols dennote Seyferts with measured $\sigma$, open ones denote Seyferts 
for which $\sigma$ has been estimated from the Faber-Jackson relation (see text).
The dashed and solid lines show the $\sigma-M_{\rm BH}$ relation of galaxies 
(Gebhardt  \et 2000a and Ferrarese \& Merritt 2000, respectively). }

\label{fig3}
\end{figure}

In figure \ref{fig3} the BH mass is plotted against the stellar velocity dispersion. 
We see that broad line AGNs are 
consistent with the $M_{\rm BH}-\sigma$ relation of inactive galaxies 
($M_{\rm BH}\propto\sigma^\alpha$, with  
$\alpha=3.5-5$; Gebhardt 
\et (2000a) find $\alpha=$3.65
and  Merritt and Ferrarese (2001a) give 4.72). 
This result is in agreement with previous results of smaller AGN samples 
(Gebhardt 2000b, Merritt \& Ferrarese 2001b) .

We can now determine whether narrow line AGNs  follow the same $M_{\rm BH}-\sigma$ relation
(which would be in odds to their lower $M_{bh}-L_{blg}$ ratios) or they lie below the broad line AGNs
in the $M_{\rm BH}-\sigma$ relation as well.
If we look carefully at the locations of the NLS1s in figure \ref{fig3} (modifiied
version of  fig 7 in Wandel 2002) 
we see that  out of the four narrow line Seyfert 1s with measured $\sigma$ 
denoted by solid pink squares in figure \ref{fig3}),
three (MRK 590, 3C120 and NGC 4051) do have 
lower Mbh than the $M_{\rm BH}-\sigma$ relation of quiescent galaxies and broad line Seyferts. 
These NLS1s (and particularely 4th, MRK 110) have  larger bulge 
luminosity than the average " $\sigma-L_{blg}$ relation" defined as the best linear to all the Seyfert galaxies with measured
bulge luminosity and velocity dispersion 
(fig. 6 of Wandel 2002). Therefore, part of the reason for the smaller $M_{\rm BH}-L_{blg}$ ratios
of NLANs could be their relatively larger bulges.

\section{Enhancing the Sample with the F-J Relation}

For AGNs without a direct velocity dispersion measurement 
 the relation between the narrow line width and the velocity
dispersion (Nelson 2000) may be used to estimate the velocity dispersion in the bulge.
Interestingly, a tight linear relation seems to exist between the virial mass derived from the Doppler width 
and the ionization parameter of  the narrow  [OIII] line 
and the BH mass in Seyfert galaxies (Wandel and Mushotzky 1986).

Here we  use the 
Faber-Jackson (F-J) relation to estimate the velocity dispersion of other AGN with measured or estimated 
bulge luminosity. 
This is useful to locate those AGNs (mainly quasars) which do not have
a measured stellar velocity dispersion. Although the F-J estimates of the velocity dispersion
are less certain since the F-J relation  has a large scatter,  it is encouraging that they
lie reasonably well within the same relation as the measured AGNs.

 The standard F-J relation is 
$L= L_o  \sigma_2^4$,
where $\sigma_2=\sigma/100\kms$ and $L_o$ is a luminosity coefficient determined
by a linear fit with a slope of 4 to the observed Seyfert Galaxies 
with measured stellar velocity dispersion.
Using this relation one may enlarge the sample and increase the dynamical range by including 
also quasars,  for a better estimate of the $\sigma -M_{\rm BH}$ relation for AGN.
We note that the NLS1s with measured $\sigma$ have a significantly larger 
luminosity coefficient than broad line ones (fig 6 of Wandel 2002):
while for broad line Seyfert 1 galaxies $L_o=10^9\ers$, for NLS1s galaxies $L_o=6\times10^9\ers$. 
We thus use the best linear fit with imposed slope of 4, $L_o=1.6\times10^9\ers$.

Doing this we see that all narrow line AGNs added in this manner (open pink squares in fig. \ref{fig3})
have lower BH masses than the value expected from the $\sigma -M_{\rm BH}$ relation.
In Wandel (2002)  a different conclusion is obtained because a larger $L_o$ is used for narrow line AGNs.
 
\section{Conclusion}

We  conclude that the lower $M_{\rm BH}/L_{bulge}$ relation of narrow line AGNs
(defined here merely by their relatively narrow H$\beta$ lines)
is a result of  a lower \rev BH mass (relatively to the broad line AGNs) - which may be reflected 
also in a lower $M_{bh}/\sigma^4$ ratio. 
Part of the effect may be due to a relatively larger $L_{bulge}/\sigma^4$ ratio 
(relative to broad line Seyferts). 

Six out of the eight NLANs with estimated $\sigma$ from the F-J relation
(open pink squares in fig. \ref{fig3}) 
 lie below the $M_{\rm BH}-\sigma$ relation (smaller BH masses).
The lower \rev BH mass of NLANs may be intrinsic or apparent, e.g. if the emission lines
of NLANs seem narrower as a result of  inclination,
if the BLR has a flattened geometry and is viewed nearly face on.

\end{document}